\documentclass[aps,prl,twocolumn,groupedaddress,amsmath,amssymb]{revtex4}

\usepackage{graphicx}% Include figure files

\usepackage{color}
\usepackage{bm}
\usepackage{latexsym}

\begin{document}

\title{Random Matrix Theory Approach to Chaotic Coherent Perfect Absorbers}

\author{Huanan Li$^1$}
\author{Suwun Suwunnarat$^1$}
\author{Ragnar Fleischmann$^{2}$}
\author{Holger Schanz$^3$}
\author{Tsampikos Kottos$^1$}
\affiliation{$^1$Department of Physics, Wesleyan University, Middletown, Connecticut 06459, USA}
\affiliation{$^2$Max Planck Institute for Dynamics and Self-organization (MPIDS),  37077 G\"{o}ttingen, Germany}
\affiliation{$^3$Institute for Mechanical Engineering, Hochschule Magdeburg-Stendal, 39114 Magdeburg, Germany}

\date{\today}

\begin{abstract}
We employ Random Matrix Theory in order to investigate coherent perfect
absorption (CPA) in lossy systems with complex internal dynamics.  The loss
strength $\gamma_{\rm CPA}$ and energy $E_{\rm CPA}$, for which a CPA occurs
are expressed in terms of the eigenmodes of the isolated cavity -- thus
carrying over the information about the chaotic nature of the target -- and
their coupling to a finite number of scattering channels. Our results are
tested against numerical calculations using complex networks of resonators and
chaotic graphs as CPA cavities.
\end{abstract}

\maketitle

{\it Introduction --} Perfect absorption is an interdisciplinary topic
relevant for a broad range of technologies, extending from acoustics
\cite{M12,MYXYS06,SBHL14,GTRMTP16} and electronic circuits
\cite{CGMM13,PCZW13,SLLREK12}, to radio frequencies (RF) \cite{DK38,S52},
microwaves \cite{S50,LSMSP08,PL10,PPVZRKL13} and optical frequencies
\cite{WLP12,CGCS10,DR12,L10,WCGNSC11,CS11,ZMZ12,STLLC14,
  PF14,KS14,VBPA15,BBFSN15,SCCWBDPCMS16}. The potential applications range
from energy conversion, photovoltaics and imaging, to time- reversal
technologies, sensing and soundproofing. In many of these applications, either
due to cost or design considerations, the requirement is to achieve maximal
absorption from minimal built-in losses. For this goal, new schemes have been
devised that exploit spatial arrangements of losses, or utilize novel
interferometric protocols. One such approach, called the Coherent Perfect
Absorber (CPA) \cite{CGCS10}, was recently proposed based on a time-reversed
laser concept.

A CPA is a {\it weakly} lossy cavity which acts as a perfect constructive
interference trap for incident radiation at a particular frequency and spatial
field distribution. This distribution is the time reversal of a lasing mode
which the cavity would emit, if the lossy medium is replaced by a gain of
equal strength. Since the outgoing signal is null due to the destructive
interferences between various pathways, the incoming waves are eventually
absorbed even in cases when a weak absorptive mechanism is employed. What
makes this approach attractive is the recent developments in wavefront shaping
of an incident wave \cite{VLN10,YPFY08,VM07}. Despite all interest, the
existing studies on CPA involve only simple cavities
\cite{M12,SBHL14,GTRMTP16,SLLREK12,S50,CGCS10,L10,WCGNSC11,CS11,ZMZ12,
  STLLC14,PF14,KS14,VBPA15,BBFSN15,SCCWBDPCMS16}. Such CPA cavities have been
realized recently in a number of experimental set- ups, ranging from optical
to RF and acoustic systems \cite{SLLREK12,WCGNSC11,GTRMTP16}.

In this paper we investigate CPA in a new set-up associated with single
chaotic cavities or complex networks of cavities coupled to the continuum with
multiple channels. The underlying complex classical dynamics of these systems
leads to complicated wave interferences that give rise to universal
statistical properties of their transport characteristics. A powerful
theoretical approach based on Random Matrix Theory (RMT) has been developed
and it has been shown that it accurately describes many aspects of such wave
chaotic systems, including the structure and statistics of spectra and
eigenstates or the distribution of transmittance, delay times, etc
\cite{RMT1,RMT2,RMT3,RMT4,RMT5,FS96,S16}.

Motivated by this success, we have used an RMT approach to derive expressions for the energy $E_{\rm CPA}$ and loss-strength 
$\gamma_{\rm CPA}$ and quantify the sensitivity of CPA on the energy and loss-strength detuning. We have also studied the statistics 
of (re-scaled) $\gamma_{\rm CPA}$, thus providing a guidance for an optimal loss-strength window for which a chaotic CPA can be 
realized with high probability. Our modeling allows the possibility of spatially non-uniform absorption which might even be localized 
at a single position. This feature is relevant for recent metamaterial proposals which advocate for the novelty of structures with spatially 
non-uniform losses (or/and gain) and can be easily realized in set-ups like the ones shown in the insets of Fig.~\ref{fig1} below. Our 
results are expressed in terms of the modes of the isolated and lossless system which contain the information about the (chaotic) dynamics. 
Specifically we find that $\gamma_{\rm CPA}$ depends on a ratio of the absolute-value-squares of eigenmode amplitudes at the boundary 
and in the absorbing regions of the system. The averaging over the statistics of eigenmode components results in non-trivial distributions 
which differ qualitatively from the well-known resonance distributions.  
We have tested the RMT results against numerical data from actual chaotic systems with non-trivial underlying dynamics, namely quantum 
graphs (Fig. 1b) \cite{KS97,KS00,GSS13,SK03}. These models of wave chaos have been already realized in the microwave regime \cite{HLBSKS12,
BYBLDS16, HBPSZZ04}. Losses can be included in a controllable manner \cite{AGBSK14} while a detailed control of the incoming waves in 
a multi-channel setting can be achieved via IQ-modulators \cite{BK16}. While in this contribution we concentrated on chaotic 
CPA traps, our approach can also serve as a basis for RMT modeling of CPA disordered diffusive cavities and CPA cavities with Anderson
localization.

%--------------------------------------------------------------------------------------
\begin{figure}[h]
\includegraphics[width=1\columnwidth,keepaspectratio,clip]{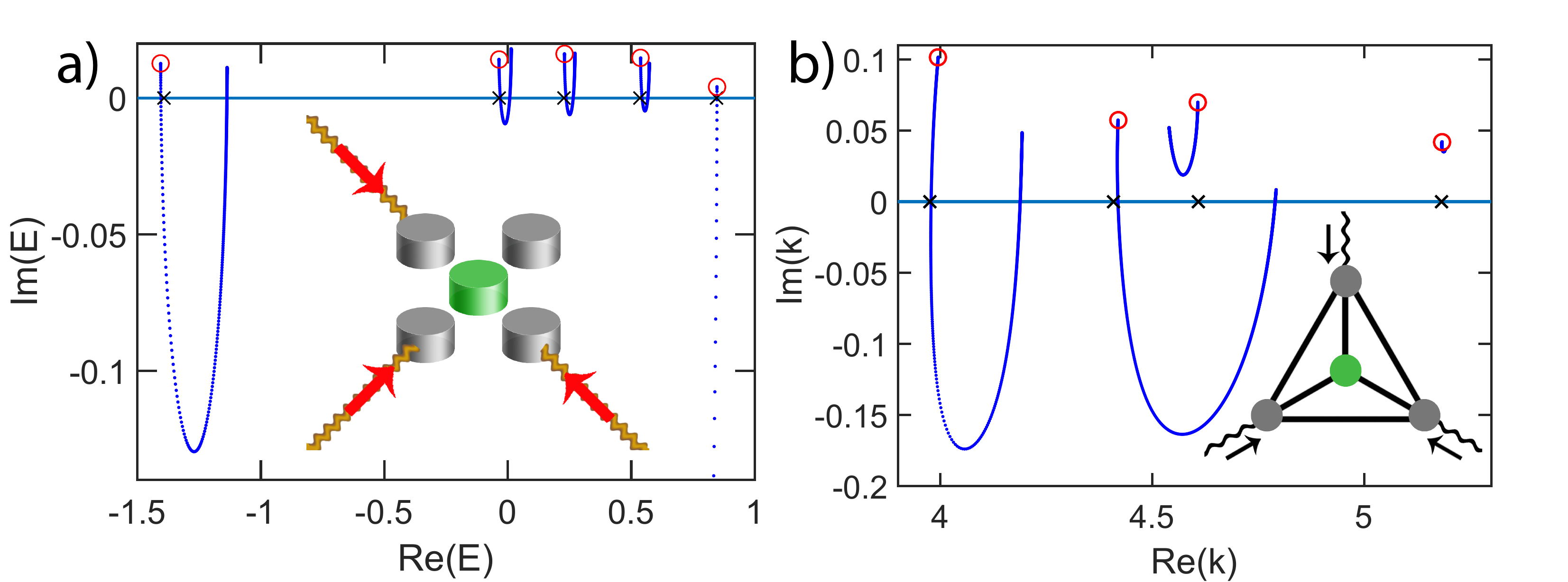}
\caption{ (color online) (a) Parametric evolution of the complex energy zeros of the scattering matrix for a random network of $N=5$ cavities 
and $M=3$ as $\gamma$ (green cavity, see inset) increases. The coupling to the leads is $w/t_L=0.15$ and the matrix elements of $H_0$ are 
taken from a box distribution $\left[-0.5,0.5\right]$. At $\gamma_{\rm CPA}$ the zeros cross the real axis i.e $\left({\cal R}e\left(E\right),{\cal I}m
\left(E\right)\right)=(E_{\rm CPA},0)$. The red circles indicate the zeros at $\gamma=0$ while the crosses indicate the eigenmodes $E_n^{(0)}$ 
of the isolated system $H_0$. (b) The same (in the complex $k$ plane) for a tetrahedron with $N=4$ vertices, $M=3$ 
leads and a lossy vertex (green, see inset) in the middle. The lossless vertices have vertex potential $\lambda {\bar l}=4\pi$. The bond lengths 
are taken from a box distribution $\left[0.5, 1.5\right]$.}
\label{fig1}
\end{figure}
%--------------------------------------------------------------------------------------

{\it RMT modeling and complex cavity networks --} The Hamiltonian that describes the isolated (i.e. in the absence of scattering channels) system is 
modeled by an ensemble of $N\times N$ random matrices $H_0$ with an appropriate symmetry: in the case that the isolated set-up is time-reversal 
invariant (TRI), $H_0$ is taken from a Gaussian Orthogonal Ensemble (GOE), while in the case of broken TRI it is taken from a Gaussian Unitary Ensemble 
(GUE) \footnote{In our simulations the matrix elements $H^0_{nm}$ (or their real and imaginary part for the GUE) are taken from a box distribution 
$\left[-W_{H_0};W_{H_0}\right]$ rather than a Gaussian distribution. In the large $N$ limit, there is no significant difference between spectrum and eigenstates 
of this ensemble and the standard GOE/GUE ensemble. The density of states of $H_0$ is semi-circle with radius $R=W_{H_0}^2 \sqrt{N}/3$ and mean 
level spacing $\Delta\sim W_{H_0}^2/\sqrt{N}$.}. Such modeling describes (in the coupled mode approximation \cite{JJWM08}) complex networks of $N$ 
coupled cavities, see inset of Fig. \ref{fig1}. The distances between the cavities are random, leading to random couplings. In this case TRI can be violated 
via magneto-optical effects. Another physical system that is described by our modeling is a set of coupled acoustic chambers or a network of random 
LC(R) elements. In the latter case the TRI can be violated via a gyrator \cite{LFLVEK14}.

Consider now that some of the cavities contain a lossy material. In this case
the Hamiltonian $H$ of the isolated system is non-Hermitian and
can be modeled as
\begin{equation}
\label{Hint}
H= H_{0}-\iota\Gamma;\qquad\Gamma=\sum_{\mu}\gamma_{\mu}\left|e_{\mu}\right\rangle \left\langle e_{\mu}\right|
\end{equation}
where $\gamma_{\mu}$ quantifies the loss in the cavity with index $\mu$ and $\{\left|e_{\mu}\right\rangle\}$ is the basis where $H_0$ is represented
(mode space). For simplicity we will assume in the following that the losses are concentrated in a single cavity $\mu=\mu_0$.

The corresponding scattering set-up is realized by coupling the isolated
system to $M$ channels that extend to infinity, see Fig. \ref{fig1}.  We
assume that these leads are one-dimensional and described by a tight-binding
Hamiltonian $H_{\rm leads}$ with matrix elements $\left(H_{\rm
  leads}\right)_{nm}=t_L\delta_{m,n\pm1}$. They support propagating waves
with a dispersion relation $E(k)=2t_L\cos(k)$, where $k$ is the
wave vector and $t_L<0$ (we set $t_L=-1$ below). We further assume that the cavities where the channels are
attached are lossless.

The scattering properties of the network are described by the $M\times M$
scattering matrix $S$ which connects incoming $|I\rangle$ to outgoing
$|O\rangle$ wave amplitudes via the relation $|O\rangle=S|I\rangle$. It can be
expressed in terms of the isolated system Eq. (\ref{Hint}) as
\begin{equation}
\label{Smatrix}
S\left(k,\gamma\right)=
-\hat{1}+2\iota\,\frac{\sin k}{t_L}W^{T}\frac{1}{H_{\rm eff}(k,\gamma)-E\left(k\right)}\,W
\end{equation}
where ${\hat 1}$ is the $M\times M$ identity matrix and $E(k)$ is the energy
of the incident plane wave \cite{data}. The rectangular $N\times M$ matrix
  $W$ contains the coupling between cavities and channels.  We assume
  $W_{nm}=w \delta_{nm}$. The effective Hamiltonian is
\begin{equation}
\label{Heff}
H_{\rm eff}\left(k,\gamma\right)= H(\gamma)+\frac{e^{\iota k}}{t_{L}}WW^{T}\,.
\end{equation}
Due to the second term the effective Hamiltonian $H_{\rm eff}$ is not
Hermitian even without internal losses i.e. $\gamma=0$.

%-----------------------------------------------------------------------------------------------------------------------------
{\it Absorption matrix and CPA conditions --} For $\gamma=0$ the
  scattering matrix is unitary, $S^{\dagger}S={\hat 1}$. For $\gamma\ne 0$,
however, this relation is violated and we introduce the operator 
$A(k,\gamma)\equiv{\hat 1}-S^{\dagger}(k,\gamma)S(k,\gamma) =A^{\dagger}$ as a
measure of the total absorption occurring in the system. For networks with one
lossy cavity, $A(k,\gamma)$ has $M-1$ degenerate zero eigenvalues while the
last eigenvalue is $0\leq\alpha(k,\gamma)\leq 1$.  When $\alpha(k,\gamma)=0$ the
  system does not absorb energy, while $\alpha(k,\gamma)=1$ indicates complete
  absorption, i.e., a CPA. This latter equation can be satisfied only for
  isolated pairs $(k_{\rm CPA}, \gamma_{\rm CPA})$ which are the object of our
  interest.  Note that unlike scattering resonances, $k_{\rm CPA}$ is real
  since it has to support a propagating incoming wave in the attached leads.

The CPA condition $\alpha(k_{\rm CPA},\gamma_{\rm CPA})=1$ is equivalent to a zero eigenvalue of the scattering
  matrix, $s_{\rm CPA}\equiv s(k_{\rm CPA},\gamma_{\rm CPA})=0$. The
  corresponding eigenvector $|I_{\rm CPA}\rangle$ identifies the shape of the
  incident field which will generate interferences that trap the wave inside
  the structure leading to its complete absorption, $\langle I_{\rm
    CPA}|S^{\dagger}(k_{\rm CPA},\gamma_{\rm CPA})S(k_{\rm CPA},\gamma_{\rm CPA})|I_{\rm CPA}\rangle=0$. 

{\it Evaluation of CPA's --} A zero eigenvalue of the scattering matrix $S(k,\gamma)$ corresponds to a pole of its inverse matrix, which can 
be represented as $S^{-1}\left(k,\gamma\right)=S(-k,\gamma)$. Thus the poles are the solutions of $\zeta(\kappa,\gamma)\equiv\det
\left(H_{\rm eff}(-\kappa,\gamma)-E(\kappa)\right)=0$, where $\kappa$ is complex in general, and the CPA's can be found numerically by 
searching for the real solutions of this equation, $\zeta(k_{\rm CPA},\gamma_{\rm CPA})=0$. Typical examples of 
the $\kappa$-evolution as $\gamma$ increases, are shown in Fig. \ref{fig1}.

We proceed with the theoretical evaluation of CPA points. We will assume that
$w\ll t_L$. Let us first consider the relevant limit of weak losses $\gamma\ll
t_{L}$. In this case $E(k,\gamma)\approx E_0(k)+\Delta E(k,\gamma)$ where
$\Delta E(k,\gamma)$ can be found via first order perturbation theory. The
unperturbed Hamiltonian is $H_0$ and we consider one particular eigenvalue
  $E^{(0)}\equiv E(k^{(0)})$ and the corresponding normalized eigenvector
  $|\Psi^{(0)}\rangle$, i.e. $H_0|\Psi^{(0)} \rangle = E^{(0)}
  |\Psi^{(0)}\rangle$. Straightforward
first order perturbation theory, together with the condition that $k_{\rm
  CPA}$ has to be real, leads to
\begin{align}
\frac{\cos\left(k^{(0)}\right)}{\cos\left(k_{\rm CPA}\right)}= & 1-\frac{1}{2}\left(\frac{w}{t_{L}}\right)^{2}\sum_{m}\left|\Psi^{(0)}_m\right|^{2}
\nonumber\\\label{CPA_perturbation}
\gamma_{\rm CPA}= & \frac{1}{2}v_{g}(k_{\rm CPA})\left(\frac{w}{t_{L}}\right)^{2}\frac{\sum_{m}\left|\Psi^{(0)}_m\right|^{2}}
{\left|\Psi^{(0)}_{\mu_0}\right|^{2}}
\end{align}
where $v_{g}(k)\equiv\frac{\partial E\left(k\right)}{\partial k}=-2t_L\sin(k)$ is the group velocity of the incoming wave while $\Psi^{(0)}_{m}$ and
$\Psi^{(0)}_{\mu_0}$ represent the components of the wave function $\left|\Psi^{(0)}\right\rangle$ at the sites $m, \mu_0$ where the leads and 
the dissipation are placed respectively. In the limiting case of $M=1$ the multichannel CPA condition Eq. (\ref{CPA_perturbation}) becomes identical 
with the critical coupling (CC) concept between input channel and loss. This is nothing else than the so-called impedance matching condition, which 
once expressed in terms of losses, indeed states that radiative and material losses must be equal \cite{H84,YY07}. At the same time, this condition is 
noticeable similar to the lasing condition, on exchanging losses with gain \cite{CGCS10}.

\begin{figure}[h]
\includegraphics[width=1\columnwidth,keepaspectratio,clip]{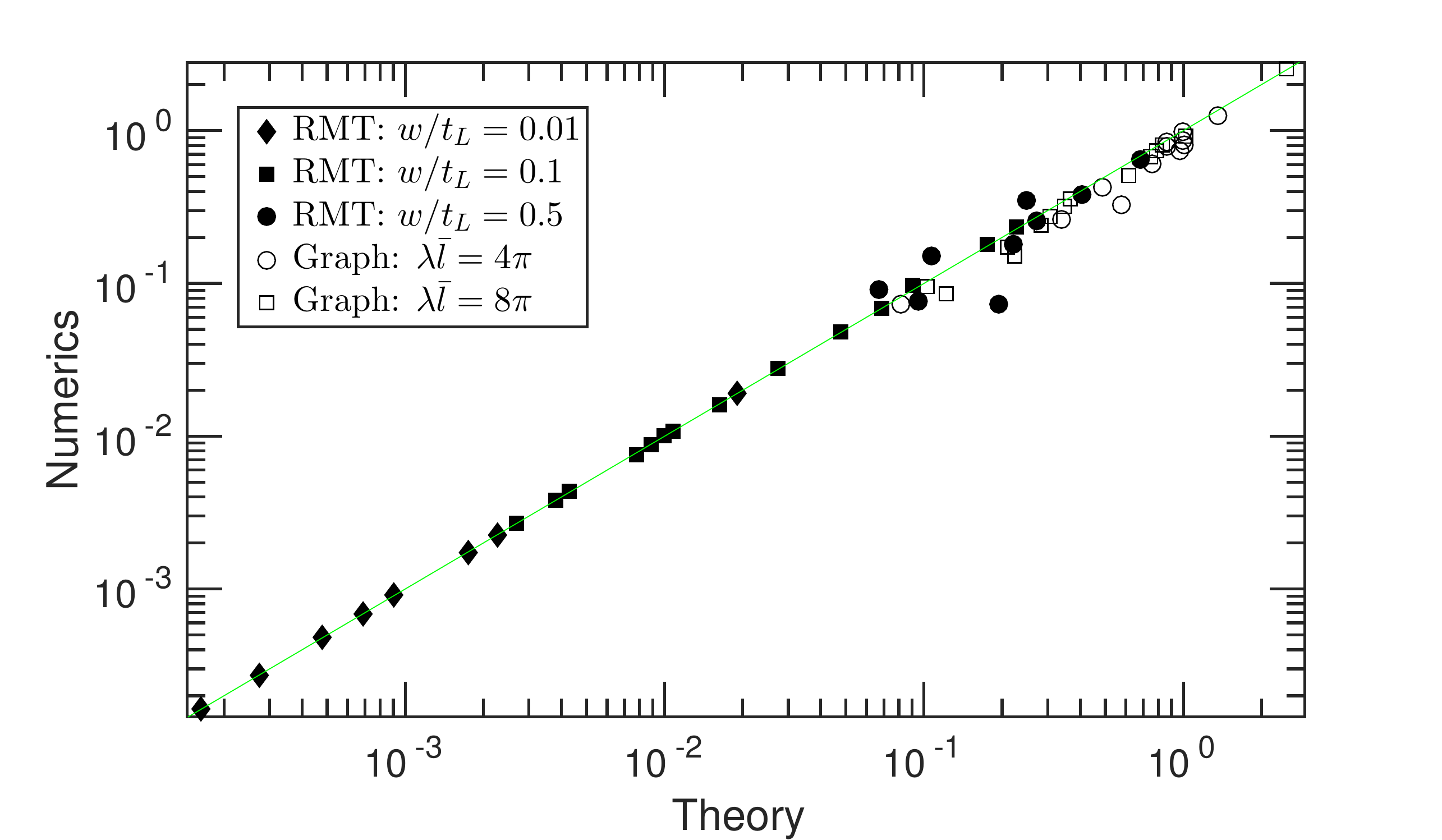}
\caption{ (color online) Theoretical versus numerical values of $\gamma_{\rm CPA}/v_{g}(k_{\rm CPA})$ for various 
realizations of a complex network of $N=15$ coupled resonators (filled symbols) and a tetrahedron graph (open 
symbols) respectively. Number of channels in both cases is $M=3$ and TRI is preserved ($\beta=1$). Various symbols correspond to different
  coupling constants $w/t_L$ for the RMT model and vertex potentials for the graph.}
\label{fig2}
\end{figure}

The accuracy of the perturbative calculation Eq. (\ref{CPA_perturbation}) is further scrutinized via direct numerical evaluations of $\left(k_{\rm CPA},
\gamma_{\rm CPA}\right)$. A comparison of the results is shown in Fig. \ref{fig2}. It is interesting to point out that although our calculations are 
applicable in the limit of weak coupling, nevertheless the agreement of the exact CPA points with the first order perturbation expressions Eq. (
\ref{CPA_perturbation}) applies for $w/t_L$ as high as 0.5.

%-----------------------------------------------------------------------------------------
%-----------------------------------------------------------------------------------------

Using Eq. (\ref{CPA_perturbation}) we can now provide a statistical description of the rescaled CPA ${\tilde \gamma}_{\rm CPA}=2\left(\frac{t_L}{w}\right)^2
\frac{\gamma_{CPA}}{v_{g}(k_{\rm CPA})}=\frac{\sum_{m}\left|\Psi^{(0)}_m\right|^{2}}{\left|\Psi^{(0)}_{\mu}\right|^{2}}$. The distribution ${\cal P} ({\tilde 
\gamma}_{\rm CPA})$ can be easily calculated using the known results for the joint probability distribution of the eigenfunction components $|\Psi^{(0)}_n|^2$ 
of a GOE (GUE) random matrix \cite{RMT1,RMT2,RMT3}. We get 
\begin{align}
\label{stat}
{\cal P}_{\beta}({\tilde \gamma}_{\rm CPA})= & {\cal N}_{\beta}
\frac{{\tilde \gamma}_{\rm CPA}^{\beta\frac{M}{2}-1}}{\left(1+{\tilde \gamma}_{\rm CPA}\right)^{\beta\frac{M+1}{2}}}
\end{align}
where $\beta=1(2)$ indicates an isolated system $H_0$ with preserved (violated) TRI. The normalization constants in front of
the above distribution are ${\cal
  N}_1=\frac{1}{\sqrt{\pi}}\frac{\Gamma\left(\frac{1+M}{2}\right)}{\Gamma\left(\frac{M}{2}\right)}$,
${\cal N}_2=M$ and $\Gamma\left(x\right)$ is the gamma function \cite{note1}.

From Eq. (\ref{stat}) we see that as the number of channels increases (the system becomes ``more open") a ``statistical gap'' is created that suppresses 
CPAs at small ${\tilde \gamma}$ (i.e. small $\gamma$ or/and large velocities $v_g$) strengths. The ``gap'' is less pronounced when $\beta=1$, since in 
this case weak localization interferences can support the trapping of the wave close to the lossy site. An estimation of the CPA gap, based on the ${\tilde 
\gamma}_{\rm CPA}$-value for which ${\cal P}({\tilde \gamma}_{\rm CPA})$ change curvature, leads to ${\tilde \gamma}_{\rm CPA} \sim 0.1 \beta M$ 
when $M\rightarrow \infty$.

A comparison of Eq. (\ref{stat}) with the numerical data for a complex network
of $N=15$ discs and $M=1,2,3$ channels is shown in Fig.~3(a). A nice
agreement is observed even though the coupling between the resonators and the
leads has a moderate high value $w/t_L\approx 0.1$.  For ${\tilde
  \gamma}_{\rm CPA}\gg 1$ \footnote{For very large values of ${\tilde
    \gamma}_{\rm CPA}$ the perturbative approach does not apply.}
the distribution Eq. (\ref{stat}) has a channel-independnet power law shape ${\cal
  P}({\tilde \gamma}_{\rm CPA})\sim 1/{\tilde \gamma}_{\rm CPA}^{{\beta\over2}+1}$. In the other
limiting case ${\tilde \gamma}_{\rm CPA}\ll 1$ we have ${\cal P}({\tilde
  \gamma}_{\rm CPA})\sim {\tilde \gamma}_{\rm CPA}^{\beta\frac{M}{2}-1}$.

%-----------------------------------------------------------------------------------------
\begin{figure}[h]
\includegraphics[width=1\columnwidth,keepaspectratio,clip]{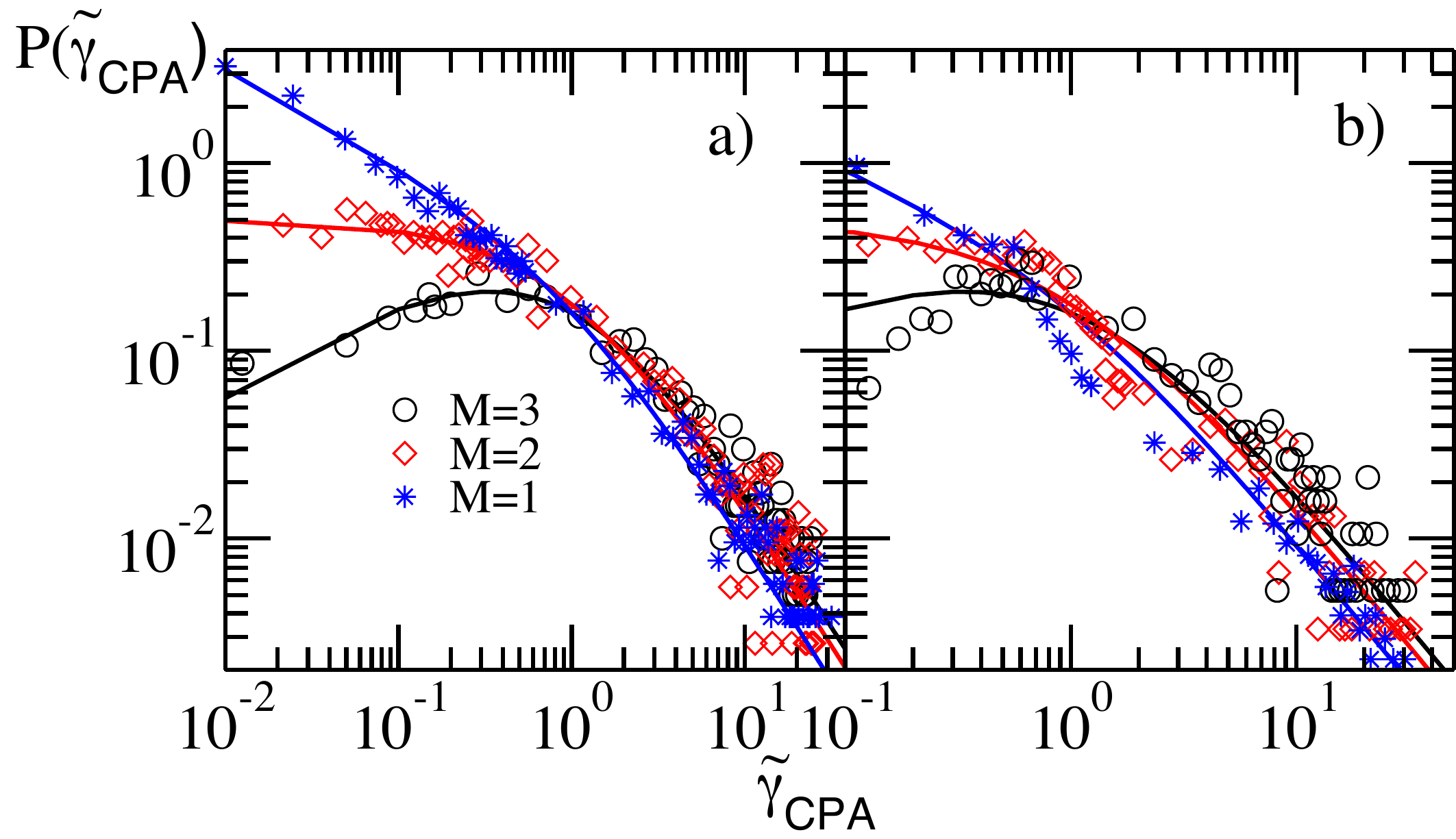}
\caption{ (color online) (a) Distribution of the rescaled CPA ${\tilde \gamma}_{\rm CPA}$ for a complex network of $N=15$ cavities 
Eqs. (\ref{Smatrix},\ref{Heff}). The numerical results are compared to the RMT prediction Eq. (\ref{stat}). (b) The same for a tetrahedron 
graph with $\lambda {\bar l}=4\pi$ at the vertices where the channels are attached. In both cases TRI is preserved, i.e. $\beta=1$.}
\label{fig3}
\end{figure}
%-----------------------------------------------------------------------------------------

Let us now calculate the incident field which can lead to a CPA. Direct substitution of Eqs. (\ref{Smatrix}), (\ref{Heff}) in the definition of $A$
allow us to re-write the absorption matrix in its spectral decomposition form:
\begin{equation}
\label{Adec}
A=-4\gamma{\sin(k)\over t_L}|\alpha\rangle\langle \alpha|;\quad |\alpha\rangle=W^{T}{1\over H_{eff}^{\dagger}-E(k)}|e_{\mu_0}\rangle.
\end{equation}
From Eq. (\ref{Adec}) it becomes obvious that the associated non-zero eigenvalue is $\alpha(k,\gamma)=-4\gamma {\sin(k) \over t_L} \langle
\alpha|\alpha\rangle$. Obviously the CPA incident waveform is given by the eigenvector $|\alpha(k_{\rm CPA},\gamma_{\rm CPA})\rangle$. 
Using again first order perturbation theory we can write the incident field $|\alpha\rangle$ Eq. (\ref{Adec}) associated with the non-degenerate 
eigenvalue $\alpha\ne 0$ of $A$ in terms of the eigenvector $\left|\Psi^{(0)}\right\rangle$ of the isolated systems $H_0$. Further substitution 
of $\left|\alpha(k_{\rm CPA},\gamma)\right\rangle$ and of $k_{\rm CPA}$ in Eq. (\ref{CPA_perturbation}) for the eigenvalue $\alpha$ gives: 
\begin{align}
\alpha(k_{\rm CPA};\gamma)= & \frac{4\gamma/\gamma_{CPA}}{(1+\gamma/\gamma_{CPA})^{2}}
\label{acpa}
\end{align}
which provides a simple expression of the relative absorption of the CPA cavity, at $k_{\rm CPA}$, when $\left|I\rangle\right. \propto
\left|\alpha(k_{\rm CPA},\gamma)\right\rangle$. Eq. (\ref{acpa}) dictates that the CPA sensitivity, defined as half-width-at-maximum, is 
proportional to $\gamma_{\rm CPA}$. Similar analysis, when $\gamma=\gamma_{\rm CPA}$, leads to the following expression 
for $\alpha(k;\gamma)$:
\begin{equation}
\label{kdef}
\alpha\left(k;\gamma_{\rm CPA}\right)=u\cdot \frac{4v_g(k)/v_g(k_{CPA})}{\left(1+v_g(k)/v_g(k_{CPA})\right)^{2}}.
\end{equation}
Note that the second term in Eq. (\ref{kdef}) has the same functional form as Eq. (\ref{acpa}). The additional factor 
$u\equiv\left[1+\frac{\cos^{2}k^{\left(0\right)}}{\left(\cos k_{CPA}-\cos k^{\left(0\right)}\right)^{2}}\tan^{2} \left(\frac{k-k_{CPA}}{2}\right)\right]^{-1}$ 
indicates that when $k_{\rm CPA}\rightarrow k^{(0)}$ the absorption eigenvalue $\alpha\left(k;\gamma_{\rm CPA}\right)$ diminishes rapidly 
as $k$ goes away from $k_{\rm CPA}$.

%-------------------

We now consider the other limit of strong losses where many of the complex solutions of $\zeta(\kappa,\gamma)=0$ turn back to the real axis 
and lead to a second CPA. The existence of CPAs for large $\gamma$ is surprising since in the over-damping domain (i.e. large $\gamma$'s) it is 
expected to have strong reflections due to impedance mismatching. However, multiple interferences in complex systems provide a  
{\it zeroes self-trapping effect} which results in the CPAs. This is analogues to the well known 
resonant self-trapping phenomenon which has been thoroughly studied in other frameworks \cite{OPR03,SZ89} and it re-surface 
also in CPAs\cite{note2}. 

%----------------------------------------------------------------------------------------------------------------------------------

{\it CPAs in Chaotic graphs --} RMT addresses universal aspects of CPAs in complex systems. At the same time one needs to be aware 
that certain non-universal features (like scarring) may emerge when CPA cavities with underlying chaotic dynamics are considered. These features 
might influence the formation of CPAs. Therefore we test our theory with a model system, where such effects are known to be prominent, namely 
quantum graphs (networks of 1D waveguides) \cite{KS97,KS00,SK03}.

A graph consists of $n=1,\cdots,V$ vertices connected by bonds. The number of bonds emanating from a vertex $n$ is its valency $v_n$ and 
the total number of directed bonds (i.e. discerning $b\equiv n\rightarrow m$ and $\bar b\equiv m\rightarrow n$) is $2B= \sum_n^V v_n$. The 
length of each bond $l_b=l_{\bar b}$ is given by a box distribution centered around some mean value ${\bar l}$ i.e. $l_b\in [{\bar l} -W_b/2,{\bar l}
+W_b/2]$. The waves on the bonds satisfy the Helmholtz equation ${d^2\Psi_b\over dx_b^2}+k^2\Psi_b=0$ (where
  $k>0$ is the wavenumber). At the vertices the wavefunction is continuous and
  satisfies the relation $\sum_{b=1}^{v_n}
  d\Psi_b/dx_b|_{x_b=0}=\lambda_{n}\Psi_b(0)$. The parameters $\lambda_{n}$ represents
  a potential concentrated on a vertex and for a lossy vertex it includes a
  negative imaginary part $-\iota\gamma_{n}$. We will restrict the losses to a
single vertex $\mu_0$ with a purely imaginary potential $-\iota\gamma$.
Leads are attached to some of the remaining vertices
$m=1,\cdots,M<V$, thus changing their valency to ${\tilde v}_m=v_m+1$. 
The details for calculating the $M\times M$
scattering matrix for this system can be found in \cite{KS00}. 
It can be represented in the form 
\begin{equation}
S=S_{MM}+S_{MB}\left({\hat 1}-{S}_{BB}\right)^{-1}\,S_{BM}\,
\end{equation}
where the $2B\times 2B$ bond-scattering matrix $S_{BB}(k,\gamma)$ describes
the multiple scattering and absorption inside the network, while the other
three blocks account for direct scattering processes and the coupling between
the leads and the network. We follow exactly the same program as for the
RMT modeling and calculate the zeros of the $S$-matrix by evaluating the poles
of its inverse $S^{-1}(k,\gamma)=S(-k,\gamma)$, i.e. by searching for the real
solutions of the secular equation $\zeta(\kappa,\gamma)\equiv\det\left({\hat
  1}-{S}_{BB} (-\kappa,\gamma)\right)=0$. Our numerical data for the case of a
fully connected tetrahedron, with one lossy vertex and $M=1,2$ and 3 leads attached to
the other vertices are shown in Fig. \ref{fig1}b and demonstrate the same
qualitative features as for the RMT model. Also the analytical evaluation
  of the CPA points via first-order perturbation theory parallels the RMT
  calculation and leads to the expression
\begin{equation}
\label{CPA_graph}
\frac{\gamma_{CPA}}{v_g(k_{\rm CPA})}= {1\over 2}\frac{\sum_{m}\left|\Psi^{(0)}_m\right|^{2}}{\left|\Psi^{(0)}_{\mu_0}\right|^{2}};\quad
k_{CPA}\approx k^{(0)}
\end{equation}
where $\Psi_{m}^{(0)}$ and $\Psi_{\mu_0}^{(0)}$ denote the values of the unperturbed wave function at the vertices with attached leads and the lossy
vertex, respectively, and the group velocity for the graphs is $v_g=\partial E/\partial k= 2k$ \cite{KS97}. The universality of this expression can be further 
appreciated by realizing its similarity with Eq.~(\ref{CPA_perturbation}) derived in the RMT framework. Eq. (\ref{CPA_graph}) has been checked 
against numerically evaluated CPA values for a tetrahedron graph, see Fig. \ref{fig2}. 

Finally we have calculated numerically the distribution of ${\tilde \gamma}_{\rm CPA}\equiv 2\gamma_{\rm CPA}/v_g(k_{\rm CPA})$ 
for graphs. The results for $M=1,2,3$ and $\lambda {\bar l}=4\pi$ are shown in Fig.~3(b) and are compared with the predictions of RMT 
Eq.~(\ref{stat}). Clearly, our theory is capable of reproducing the main features of the distribution also for this model, where prominent scarring effects 
are known to exist \cite{GSS13,SK03}. However, the agreement with Eq. (\ref{stat}) is less convincing for strong coupling to the leads ($\lambda=0$, not 
shown) where a non-perturbative approach is necessary. 

{\it Conclusions -} We investigated the distribution of loss-strengths for the realization of a chaotic CPA using a RMT formalism. In the case of $d$ 
absorbers one gets 
a distribution with power law tails that might even (e.g. for $d=1$) not possess a mean value. This has to be contrasted to the case of uniform losses where 
one ends up with a $\chi^2$- distribution with exponentially decaying tails and well-defined mean. Furthermore for non-uniform losses we have discover 
the novel effect of {\it zeroes self-trapping} which is absent for uniform losses. Finally we evaluated the robustness of CPA with respect to loss-detuning and
we found that in case of frequency-detuning the upper bound for CPA robustness is controlled by the mean level spacing. The effects of semiclassical features, 
like scarring etc, are a subject of ongoing research.

\textit{Acknowledgement - }  (H.L, S.S. and T.K) acknowledge support from an AFOSR MURI grant FA9550-14-1-0037.
%%%%%%%%%%%%%%%%%%%%%%%%%%%%%%%%%%%%%%%%%%%%%%%%%%%%%%%%%%%%%%%%%%%%%%%%%%%%%%%%%%%%

%%%%%%%%%% Merge with supplemental materials %%%%%%%%%%
\pagebreak
\widetext
\begin{center}
\textbf{\large Supplemental Materials}
\end{center}

%%%%%%%%%% Merge with supplemental materials %%%%%%%%%%
%%%%%%%%%% Prefix a "S" to all equations, figures, tables and reset the counter %%%%%%%%%%
\setcounter{equation}{0}
\setcounter{figure}{0}
\setcounter{table}{0}
\setcounter{page}{1}
\makeatletter
\renewcommand{\theequation}{S\arabic{equation}}
\renewcommand{\thefigure}{S\arabic{figure}}
\renewcommand{\bibnumfmt}[1]{[S#1]}
\renewcommand{\citenumfont}[1]{S#1}
%%%%%%%%%% Prefix a "S" to all equations, figures, tables and reset the counter %%%%%%%%%%

We have checked numerically the validity of the theoretical predictions Eqs. (7,8) of the main text about the sensitivity of the CPA to 
wavenumber $k$ and loss-strength $\gamma$ detuning.  To this end we have used an RMT model with one absorber and evaluated 
the eigenvalues of the absorption matrix. In this case there is at most one non-zero eigenvalue $\alpha$. In Fig. S1 we report the behavior 
of $\alpha$ versus $k$ and $\gamma$ for a system of two channels $M=2$. The peak indicates the position of $k_{\rm CPA},\gamma_{\rm CPA}$.
The blue continuous and black dashed lines indicate the theoretical predictions of Eqs. (7) and (8) respectively. 

%-----------------------------------------------------------------------------------------
\begin{figure}[h]
\includegraphics[width=0.8\columnwidth,keepaspectratio,clip]{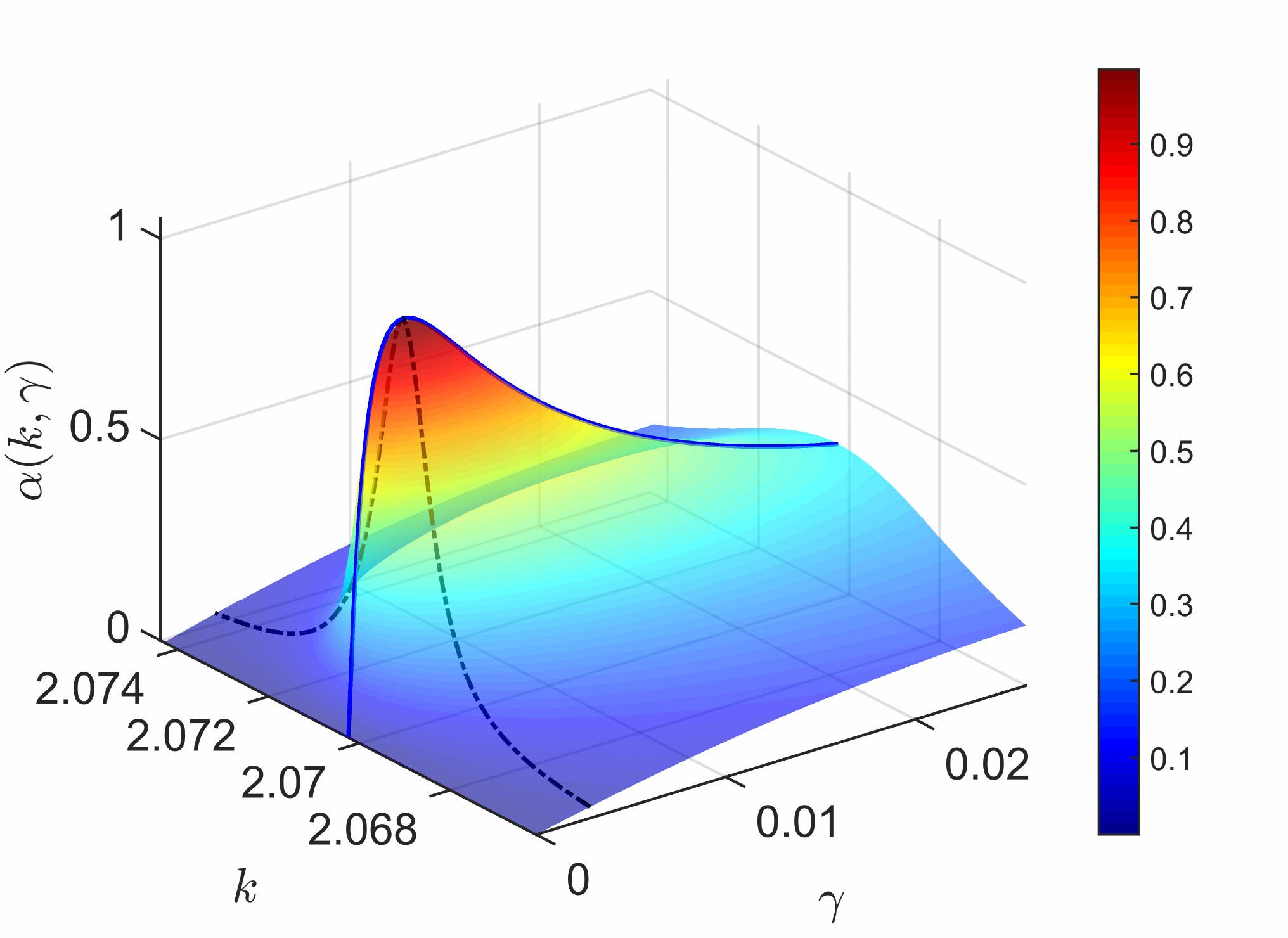}
\caption{ (color online) The non-zero absorption eigenvalue  $\alpha(k,\gamma)$ versus the wavenumber $k$ and the loss-strength  $\gamma$ 
for a RMT system of  $N=5$ sites and two channels, with one lossy site in the middle (see inset of Fig. 1). The blue continuous and black dashed   
lines indicate the theoretical predictions of Eqs. (7) and (8) respectively.
}
\label{fig3}
\end{figure}
%-----------------------------------------------------------------------------------------


\begin{thebibliography}{99}
\bibitem{note1} Eq. (\ref{stat}) can be extended to the case of $d$ equal absorbers (none of which are placed at the
resonators coupled to the leads) leading to ${\cal P}_{\beta}({\tilde \gamma}_{\rm CPA}) = \mathcal{\tilde{N}}_{\beta}\frac{\tilde{\gamma}_{CPA}^{\beta\frac{M}{2}-1}}
{\left(1+\tilde{\gamma}_{CPA}\right)^{\beta\frac{M+d}{2}}}$ where $\mathcal{\tilde{N}}_{\beta}= \frac{\Gamma\left(\beta\frac{M+d}{2}\right)}
{\Gamma\left(\beta\frac{d}{2}\right)\Gamma\left(\beta\frac{M}{2}\right)}$.
\bibitem{note2} The applicability of RMT to the ``super-radiance" CPA domain implies that other similar (in spirit) effects like super-resolution 
etc might also benefit from such approach.
\end{thebibliography}
\end{document}